\documentclass[a4paper,10pt]{article}

\usepackage{amsmath,amssymb,mathrsfs,bm}
\usepackage{calrsfs}
\usepackage{color,graphicx}
\usepackage{authblk}
\usepackage[left=2.5cm,right=2.5cm,top=3cm]{geometry}
\newcommand{\bA}{\mathbf{A}}

\newcommand{\bB}{\mathbf{B}}
\newcommand{\bb}{\mathbf{b}}
\newcommand{\bBe}{\mathbf{B}_{\mathrm{e}}}

\newcommand{\bD}{\mathbf{D}}
\newcommand{\bbD}{\mathbb{D}}
\newcommand{\bE}{\mathbf{E}}

\newcommand{\be}{\mathbf{e}}
\newcommand{\bF}{\mathbf{F}}
\newcommand{\FF}{\mathcal{F}}
\newcommand{\bFe}{\mathbf{F}_{\mathrm{e}}}

\newcommand{\bG}{\mathbf{G}}
\newcommand{\bh}{\mathbf{h}}
\newcommand{\bH}{\mathbf{H}}
\newcommand{\bI}{\mathbf{I}}

\newcommand{\bL}{\mathbf{L}}
\newcommand{\bPsi}{\bL}

\newcommand{\bn}{\mathbf{n}}

\newcommand{\bN}{\mathbf{N}}

\newcommand{\bQ}{\mathbf{Q}}

\newcommand{\bt}{\mathbf{t}}

\newcommand{\bT}{\mathbf{T}}
\newcommand{\bTa}{\mathbf{T}_{\text{a}}}

\newcommand{\bW}{\mathbf{W}}

\newcommand{\bv}{\mathbf{v}}
\newcommand{\bx}{\mathbf{x}}
\newcommand{\bX}{\mathbf{X}}

\newcommand{\bQd}{\bQ^{\Diamond}}

\newcommand{\eps}{\varepsilon}
\newcommand{\si}{\sigma}
\newcommand{\sio}{\sigma_{0}}
\newcommand{\uctd}{\triangledown}

\newcommand{\Ldot}[1]{\overset{\bm{.}}{#1}}

\newcommand{\Pt}{\mathcal{P}_t}
\newcommand{\bnu}{\bm{\nu}}
\newcommand{\Wext}{W^{(\text{ext})}}
\newcommand{\spr}{S_{\mathrm{pr}}}
\newcommand{\sa}{\sigma_{\text{active}}}

\DeclareMathOperator{\tp}{\otimes}

\DeclareMathOperator{\tr}{tr}
\DeclareMathOperator{\er}{e}

\DeclareMathOperator{\divr}{div}

\newcommand{\D}[2]{\frac{\partial #1}{\partial #2}}

\newcommand{\str}[1]{#1^{\star}}
\newcommand{\comm}[1]{{\color{black}#1}}
\newtheorem{lemma}{\bf Lemma}[section]

\title{Active viscoelastic nematics with partial degree of order}
\author{Stefano Turzi \thanks{\texttt{stefano.turzi@polimi.it}}}

\affil{Dipartimento di Matematica, Politecnico di Milano, Piazza Leonardo da Vinci 32, 20133 Milano, Italy}

\date{\today}

\begin{document}
\maketitle

\begin{abstract} Continuum models of active nematic gels have proved successful to describe a number of 
biological systems consisting of a population of rodlike motile subunits in a fluid environment. However, in 
order to get a thorough understanding of the collective processes underlying the behaviour of active biosystems, the 
theoretical underpinnings of these models still need to be critically examined. To this end, we derive a minimal 
model based on a nematic elastomer energy, where the key parameters have a simple physical interpretation and the 
irreversible nature of activity emerges clearly. The interplay between viscoelastic material response and active 
dynamics of the microscopic constituents is accounted for by material remodelling. Partial degree of order and defect dynamics is included as a 
result of the kinematic coupling between the nematic elastomer shape-tensor and the orientational ordering tensor $\bQ$. In a simple one-dimensional channel geometry, we use linear stability analysis to show that even in the isotropic phase the interaction between 
flow-induced local nematic order and activity results in a spontaneous flow of particles.
%In a simple one-dimensional channel geometry, we use linear stability analysis to show that the interaction between 
%nonuniform nematic order and activity results in a spontaneous flow of particles even in the absence of a preferred 
%nematic phase.
\end{abstract}

\section{Introduction}
\label{sec:intro}
Active nematic gels are fluids or suspensions of orientable rodlike objects endowed with active dynamics. They are a 
recent physical paradigm used to represent the collective dynamics of many biological systems such as filaments of the 
cytoskeleton \cite{13Sanchez,15Piel}, dense bacterial suspensions \cite{08Volfson,13Wioland,16Wioland}, and, on a more 
macroscopic 
scale, also flocking of birds or fishes \cite{14Cavagna}. These systems are intrinsically out of equilibrium, because 
the particles continuously consume energy that is used for their active movement or to exert mechanical forces. The 
interplay between many of these active particles can lead to very complex patterns of collective motion and 
self-organized structures \cite{09edwards,12Giomi,13Marchetti,15Giomi,19Li,20Turzi}, with features not observed in passive 
systems, such as internally generated flow patterns, large-scale collective motion, active turbulence, and 
sustained oscillations.

These systems have very different microscopic interactions among their constituents, however they share a 
number of qualitative features in their collective dynamics. This implies that it is possible to provide a macroscopic 
description that overlooks the fine microscopic details and it is based only on very general principles such as 
compatibility with thermodynamics and the symmetry properties induced by the material symmetry of the 
microscopic constituents. However, these general principles still leave a lot of freedom in formulating a model and 
the relative importance of the many phenomenological parameters introduced have to be determined from experiments, 
which are difficult to perform and interpret \cite{19Li,21Vitelli}. 

To this end, it is important to carefully formulate simple models in order to advance our understanding of how the 
physics of active matter is relevant in biological contexts. Continuum models based on nematic liquid crystal dynamics, 
namely Ericksen-Leslie theory if the orientational order is described by the director $\bn$ or Beris-Edwards theory if 
the ordering tensor $\bQ$ is employed, seem to have been particularly successful in describing the rich physics of 
active matter 
\cite{05julicher,07julicher,07MarenduzzoPRL,09edwards,11Cates,14Giomi,15Prost,15Cates,15Yeomans,17Yeomans}, at least 
from a qualitative point of view. These active models typically add two key ingredients to the classic hydrodynamic 
equations of liquid crystals: a non-equilibrium stress term owing to activity, and a Maxwell relaxation time to account 
for the viscoelastic behaviour of most biological fluids. 

The active contribution to the Cauchy stress tensor accounts for the chemical 
energy consumed by the microscopic constituents to generate macroscopic motion and is generally chosen to be 
proportional to the nematic ordering tensor, $\sa = -\zeta \bQ$, where the sign of $\zeta$ depends on whether
the active particles generate contractile or extensile stresses \cite{02Rama, 04Rama}. Active nematics are 
characterized by a strong deviation from thermal equilibrium due to the environmental energy supply 
and it is possible to observe complex dynamics, defects formation, and active-driven turbulence 
\cite{06Voituriez,11Cates,14Giomi,17Shendruk}. Therefore, to investigate dynamic instabilities, defect formation, and defect 
motion, it is important to use de Gennes' $\bQ$-tensor as an order descriptor in the model. We will consider for 
simplicity a dense single-phase material, although multiphase models have also been developed to 
describe cell separation mechanisms \cite{12Cates,14GiomiDS,15Yeomans}. \\

However, despite the many theoretical and numerical studies, the introduction of the active term, $\sa$, to the stress 
tensor, although attractive for its simplicity, is not completely satisfactory from a physical standpoint and there are 
some critical points that can be raised against it. A first remark is related to the compatibility of $\sa$ with 
irreversible thermodynamics \cite{14Brand}. It is possible to use classical irreversible thermodynamics to derive a 
thermodynamically consistent coupled chemo-mechanical theory of active nematics, see for example 
Refs.\cite{05julicher,16Forest}. However, $\sa$ appears in these models as a reactive, i.e., a time-reversible term.  
From a physical perspective, this implies that the transfer of chemical energy into macroscopic motion is a 
reversible process, so that it could be possible, in principle, to use particle locomotion to generate chemical fuel. 
However, it is natural to assume that the motile subunits could only consume chemical energy, so that activity should 
be introduced as an irreversible process.

It is interesting to note that a term of the type $\sa$, proportional to $\bQ$, is indeed thermodynamically correct at 
first order. More precisely, it is possible to put forth a thermodynamically consistent theory of active nematics based 
on nematic elastomer energy and microscopic relaxation dynamics \cite{17tur}, where activity is represented as an 
external remodelling force, which clearly distinguishes it from the commonly used time-reversible 
representation. When this latter theory is approximated under the assumption of fast material remodelling times and 
low activity, the first order contribution to the stress tensor turns out to be proportional to $\bQ$, exactly as in 
the classical theories inspired by liquid crystal physics.

A second critical point is that the active stress power, $\sa \cdot \nabla \bv$, vanishes in the absence of 
macroscopic motion. In turn, this implies that there is no chemical fuel consumption when $\bv=0$. However, chemical 
fuel (e.g., ATP molecules) can also be used for mesoscopic motion such as material reorganization at the microscale, 
or for polymer stiffening, without any visible macroscopic velocity \cite{17tur,19Turzi,19Bacca}. The transduction of 
chemical energy into mechanical work is thus internal to the material and is more related to the evolution of the 
physical cross-links and their reorganization rather than to the generation of macroscopic flow.

%Finally, while $\bTa$ certainly possess the correct symmetry, its justification based on microscopic arguments is not 
%entirely satisfactory \cite{02Rama, 04Rama}. The original motivation rests on the multiple expansion of the Stokes 
%flow 
%field around a single swimmer, a far-ﬁeld expansion for the velocity ﬁeld around a force distribution 
%\cite{17Yeomans}. 
%For active systems the contribution to the continuum stress tensor comes from the dipole terms in the multipole 
%expansion. However, this derivation applies to a single swimmer that moves autonomously, when the velocity field is 
%described far from the swimmer. How is it possible that this derivation also apply to dense suspensions of particles 
%where the ``swimmers'' are closely packed and their interaction is mainly steric (think for example to dense bacterial 
%colonies or dense actomyosin filaments)?

Viscoelastic effects are important to capture the types of slow dynamics that one might expect, e.g., in 
the cytoskeleton, which contains long-chain flexible polymers and other cytoplasmic components that have long intrinsic 
relaxation times. This feature is common to many biological tissues, but the standard continuum 
models of active matter, derived from liquid crystal physics, fail to capture this slow viscoelastic dynamics as assume 
short local relaxation times. This is a major shortcoming because viscoelastic effects are expected to couple strongly 
with the active liquid-crystalline dynamics and thereby potentially radically modify the effects of 
activity. The interplay between viscoelasticity and active motion have been explored in a number of recent papers 
\cite{15Cates,16Cates,20Yeomans}. While the models in Refs.\cite{15Cates,16Cates,20Yeomans} introduce viscoelasticity 
by adding new terms to the free energy that account for the viscoelastic polymer and couple the nematic tensor $\bQ$ 
with the polymeric conformation tensor, we here propose a theory that kinematically links the nematic-elastomer shape 
tensor (or step-length tensor), $\bL$, with the ordering tensor, $\bQ$. In so doing, viscoelastic features come out 
naturally from an 
active nematic elastomer model endowed with material relaxation. This latter feature is introduced by using a multiplicative 
decomposition of the deformation gradient, a classical technique in plasticity theory \cite{92SimoA,92SimoB,98Reese}. 
Similar ideas have been recently applied to describe viscoelastic soft solids with reinforcing fibres 
\cite{21Ciambella}. 

Starting from a classical nematic elastomer free energy, and adding material relaxation, active 
remodelling tensor and linking the shape tensor to the nematic ordering tensor, we are able to propose a rather simple 
theory 
that is constructed on the basis of rational thermodynamics and accounts for defects formation, dynamics and 
viscoelastic effects, all of which seem to be essential ingredients for a sound description of active biological matter 
at the continuum scale.

The paper is organised as follows. The model is presented in \S\ref{sec:model}. In \S\ref{sec:fast_relaxation} we assume fast relaxation times and develop a fluid-like approximation that allows us to make a comparison with the existing theories. A linear stability analysis is performed in \S\ref{sec:fast_relaxation}\ref{sec:applications} to show that a spontaneous flow may arise even in the absence of any Landau-de Gennes potential that favours alignment. \comm{The opposite limit of nearly elastic behaviour is briefly explored in \S\ref{sec:elastic}, where we show that activity can induce actuation in nematic elastomers.} The conclusions are drawn in \S\ref{sec:conclusions}. Some technical derivations are given in the appendices for ease of reading.
%\end{fmtext}

\section{Theory}
\label{sec:model}

After De Gennes introduced the ordering tensor $\bQ$ to describe partial order in nematics, many authors have proposed general continuum theories of nematics with tensorial order. Among these we would like to mention the works of Pereira Borgmeyer and Hess \cite{95Hess}, Qian Sheng \cite{98Qian}, Beris Edwards \cite{94Beris}, within the framework of irreversible thermodynamics, Sonnet Virga \cite{04Virga}, who employ a new variational principle, and Stark Lubensky \cite{03Lubensky}, who use a general Poisson-bracket formalism. A critical account of these theories can be found in Refs.\cite{04Virga,12Virga_book}. 

Here we follow an alternative, and simpler, route by considering a theory based on a nematic elastomer free energy, material remodelling, kinematic (instead of energetic) coupling of the elastomer shape tensor and the nematic ordering tensor, and active external remodelling forcing. We try to present the theory in its simplest form, i.e., whenever possible we choose the least number of phenomenological parameters (in the same spirit as the one constant approximation for Frank's elastic energy in liquid crystal theory). However, at some places we will indicate where a simple extension of the theory is possible.

In order to enforce a deformation of an elastic solid, like an elastomer, we must apply a load. As soon as the external 
forces are released, the material returns to its natural, unstressed configuration. At a microscopic level, during such 
process, the positions of the strained molecules are distorted without significant variations of their relative 
arrangement. By contrast, when a deformation is applied to a fluid, a further degree of freedom comes into play: 
molecular reorganization. Indeed, by suitably modifying the relative positions between material points, the system
proves able to reduce and even eventually cancel the existing stresses just by adapting the natural configuration.
This material remodelling, consisting basically in a relabelling of which molecules are the first neighbours of which, 
turns out to be able to reduce the stresses without making resort to any collective molecular motion. In other words, 
the relaxation process drives the natural, unstressed configuration closer to the present configuration. Nevertheless, 
experimental evidence suggests that not all the strains may be recovered by simply reorganizing the natural 
configuration. In particular, fluids are not able to relax density variations, as each fluid possesses a reference 
density dictated by the microscopic fact that each molecule occupies in average a well-defined volume. As a 
consequence, a necessary feature of a physically meaningful model of an accommodating fluid is that the energy cost of 
any density variation should not be compensated by the microscopic relaxation. 

\begin{figure}[!h]
\begin{center}
\includegraphics[width=0.5\linewidth]{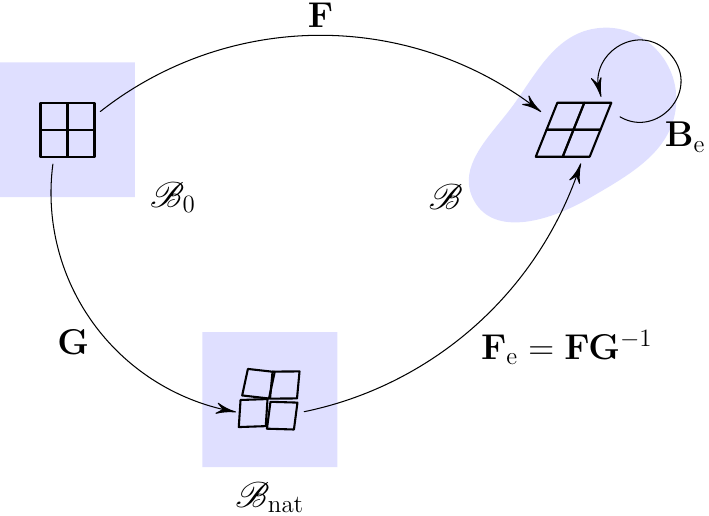}
\caption{Multiplicative decomposition of the deformation gradient $\bF$ in effective component $\bFe$ and remodelling 
component $\bG$. $\mathcal{B}_{0}$, $\mathcal{B}_{\mathrm{nat}}$ and $\mathcal{B}$ are, respectively, the reference, 
the 
natural and the current configurations.  The effective left Cauchy-Green deformation tensor, $\bBe$, is a \comm{spatial tensor field, maps the 
tangent space to the current configuration at $\bx$ in itself, i.e., spatial vectors to spatial vectors.}}
\end{center}
\label{fig:01}
\end{figure}

A now standard way to introduce material remodelling in a continuum theory is to use the Kr\"{o}ner-Lee-Rodriguez 
multiplicative decomposition \cite{59Kroner,69Lee,94Rodriguez} for the deformation gradient $\bF=\bFe \bG$. The 
effective deformation tensor, $\bFe$, will measure the elastic response, from an evolving natural configuration, 
$\mathcal{B}_{\mathrm{nat}}$ (Fig.\ref{fig:01}). An alternative similar approach, which introduces metric degrees 
of freedom, has been recently used to describe active chiral viscoelastic materials \cite{22Lier}. The relaxing 
deformation tensor, $\bG$, determines how this 
configuration locally departs from the reference configuration. Since the elastic response is determined by $\bFe$, 
only the effective deformation appears explicitly in the strain energy. By contrast, energy dissipation (entropy 
production) is only associated with the evolution of $\bG$. The same multiplicative decomposition is now a standard 
tool in continuum theories to describe plastic behaviour or growth phenomena. It has also been recently applied to 
explain the hints of viscoelasticity that remain at the hydrodynamic level when a sound wave propagates inside a 
nematic crystal \cite{14bdt,15tur,16bdt,16Turzi}.

For later convenience we also define the inverse relaxing strain $\bH = (\bG^{T}\bG)^{-1}$, so that the 
\emph{effective} left-Cauchy-Green deformation tensor can be written as
\begin{equation}
\bBe = \bFe\bFe^{T} =\bF\bG^{-1} \bG^{-T}\bF^{T} = \bF \bH \bF^{T}.
\end{equation}

For the sake of simplicity, we assume the standard energy of nematic elastomers, where the uniaxial symmetry of the 
constituents is reflected in the uniaxial symmetry of both the shape tensor $\bL$ and the ordering tensor $\bQ$. The 
free energy density per unit mass is taken to be, 
\begin{equation}
	\sigma(\rho,\bBe,\bL,\bQ,\nabla\bQ) = \sio(\rho) + \tfrac{1}{2}\mu \Big(\tr \big(\bPsi^{-1}\bBe - \bI \big) - 
	\log\det\big(\bPsi^{-1}\bBe \big)\Big)
	+ \sigma_{\text{LdG}}(\bQ) + \sigma_{\text{el}}(\bQ,\nabla\bQ),
	\label{eq:DensitaEnergiaPsiQ}
\end{equation}
which comprises both the elastomer elastic energy, written in terms of $\bL$, and the nematic energy, written in terms 
of $\bQ$. At this stage $\bL$ and $\bQ$ are still independent, but they will be kinematically related in the following. 
The first term in Eq.\eqref{eq:DensitaEnergiaPsiQ},  $\sio(\rho)$, is a volumetric term that depends only on the 
density 
$\rho$ and does not relax. By contrast, the second term, namely the neo-Hookean nematic elastomer energy density 
\cite{Warner}, 
with shear modulus $\rho \mu$, depends here only on the effective deformation tensor $\bBe$ and this is a signature 
that it undergoes a remodelling dynamics (which will be described by an equation for $\bBe$). 
The third and fourth terms, $\sigma_{\text{LdG}}(\bQ)$ and $\sigma_{\text{el}}(\bQ,\nabla\bQ)$, 
are the standard Landau-de Gennes thermodynamic potential and nematic elastic energy. We will use the classical 
expressions
\begin{align}
\sigma_{\text{LdG}}(\bQ) & = \tfrac{1}{2} A \tr(\bQ^2) - \tfrac{1}{3} B \tr(\bQ^3) + \tfrac{1}{4} C \tr(\bQ^2)^2, 
\label{eq:LdG}\\
\sigma_{\text{el}}(\bQ,\nabla\bQ) & = \tfrac{1}{2}k \| \nabla \bQ \|^2, \label{eq:Frank}
\end{align}
where, however, more complicated formulas can be used if, for instance, it is important to use different elastic 
constants. 
In terms of the director, $\bn$, and the degree of order, $S$, the tensors $\bL$ and $\bQ$ are typically written as 
\begin{align}
\bPsi & = a(S)^2 (\bn \tp \bn) + a(S)^{-1}\big(\bI - \bn \tp \bn \big), \label{eq:shape-tensor}\\
\bQ & = \tfrac{2}{3}S (\bn \tp \bn) - \tfrac{1}{3}S\big(\bI - \bn \tp \bn \big), \label{eq:Quniaxial}
\end{align}
where $a$ is the material parameter\footnote{In nematic elastomer theory the shape parameter $a$ is usually denoted 
with $\ell_{\parallel}$, and called effective step-length along the direction parallel to $\bn$. However, we prefer to 
use $a$ since our theory, in the limit of fast relaxation times, also applies to liquid crystals where the concept of 
step-length, which is closely related to polymer physics, is not clearly defined.} that gives 
the amount of spontaneous elongation along $\bn$ in an uniaxially ordered phase. The shape tensor is spherical, prolate, 
or oblate, respectively, for $a=1$, $a>1$ and $a<1$. The tensor $\bL$ is symmetric and satisfies $\det\bL = 1$. By 
contrast, $\bQ$ is symmetric and traceless. 

It must be noted that in our theory we do not assume $\bL$ and $\bQ$ as independent 
observables, so that Eq.\eqref{eq:shape-tensor} is not assumed to be valid 
\emph{a-priori}. We rather express $\bL$ as a function of $\bQ$. One simple strategy to relate the shape tensor $\bL$ 
with the ordering tensor $\bQ$ is to write \cite{97Biscari,14Keip,15Calderer,Warner} $\bQ = f(\bL) = \bL - 
\frac{1}{3}\tr(\bL)\,\bI$, so that $\bQ$ is automatically traceless by construction. Despite its simplicity, this 
equation is not completely satisfactory. First, 
we would like to express $\bL$ as a function of $\bQ$, but the inversion of the previous formula is not 
straightforward. 
More precisely, $\bL$ is not uniquely defined by the inversion of $\bQ = f(\bL)$, for a given $\bQ$. It is uniquely 
defined 
only once we impose the additional condition $\det\bL = 1$, but the unit determinant property is not a consequence of 
the functional relation between $\bQ$ and $\bL$, i.e, it does not follow automatically from the properties of $\bQ$. 
Furthermore, since $S$ ranges in $[-\tfrac{1}{2},1]$, the maximum achievable value of the shape parameter is 
$a_{\text{max}} \approx 1.32$. This means that we need to introduce an additional material parameter, since otherwise 
it would not be possible to represent even moderate anisotropic situations, in contrast to the estimated value $a 
\approx 2$ found in \cite{14bdt,16Turzi}. 

There is, however, a unique functional relationship linking $\bL$ to $\bQ$ that verifies $\tr(\bQ)=0$ if and only if 
$\det(\bL)=1$, and this is the exponential matrix. Therefore, it is natural to posit
\begin{equation}
\bL = \er^{b \bQ},
\label{eq:LexpQ}
\end{equation}
where $b$ is a material parameter that measures how much the nematic order given by $\bQ$ affects the shape tensor 
$\bL$. In particular, we find that $a(S) = \er^{b S/3}$.

We base the derivation of the equations of motion on the free energy imbalance for mechanical theories \cite{Gurtin,Rubin}: 
the power expended by the external forces on a convecting spatial region $\Pt$ must be greater than or equal to the 
temporal increase in kinetic and free energy of $\Pt$. The difference being the power dissipated in irreversible 
processes. Specifically, for any isothermal process, for any portion $\Pt$ 
of the body at all times, we require
\begin{align}
\mathcal{D} := \Wext - \dot{K} - \dot{\FF} \geq 0 , 
\end{align}
where $\Wext$ is the power expended by the external forces, $\dot{K}$ is the rate of change of the kinetic energy, 
$\dot{\FF}$ is the rate of change of the free energy, and the dissipation $\mathcal{D}$ is a positive quantity that 
represents the energy loss due to irreversible processes (entropy production). Here, an overdot indicates the material 
time derivative. Furthermore, we assume that no positive dissipation is associated with a rigid rotation of the whole
body. A final key assumption is that positive dissipation is only associated to the evolution of
the natural or stress-free configuration of the body, i.e., entropy is produced only when microscopic reorganization 
occurs.

The equations of motions can now be derived following the thermodynamic procedure outlined in \cite{17tur}. The details 
of this derivation when the tensor $\bQ$ is employed are reported in Appendix \ref{app:equations}. We simply report 
the main equations here. The dissipation is derived in Appendix \ref{app:equations}, see 
Eqs.\eqref{eq:dissipazioneIntegrale}, \eqref{eq:dissipazioneIntegrale2}, and it is reported here below for ease of 
reading
\begin{align}
\mathcal{D} & = \int_{\Pt} \left(\bb - \rho \dot{\bv} + \divr\bT \right) \cdot \bv \,\,dv 
+ \int_{\partial\Pt} \left(\bt_{(\bnu)} - \bT\bnu \right) \cdot \bv \,\,da 
- \int_{\Pt} \bh \cdot \Ldot{\bQ} \,\,dv \notag \\
& - \int_{\partial\Pt} \left(\rho \D{\si}{\nabla \bQ}\bnu \right) \cdot \Ldot{\bQ}\, da
+ \int_{\Pt}\left(\bTa-\rho\D{\si}{\bBe}\right)\cdot \bBe^{\uctd}\,\, dv,
\label{eq:dissipazioneIntegrale_text}
\end{align}
where $\bb$ is the external body force, $\bT$ is the Cauchy stress tensor (as defined in \eqref{eq:defT}), $\bh$ is 
the molecular field (as defined in \eqref{eq:defh}), and $\bTa$ is the activity tensor. The codeformational 
time-derivative is defined as $\bBe^{\uctd} := 
(\bBe)\Ldot{\phantom{i}} - (\nabla\bv) \,\bBe - \bBe 
\,(\nabla\bv)^T$. \comm{Since $\bBe(t,\bx)$ is a \emph{spatial tensor field}, the Cartesian components of $\bBe^{\uctd}$ are explicitly calculated as
\begin{equation}
\big[\bBe^{\uctd}\big]_{ij} = \D{[\bBe]_{ij}}{t} 
+ \sum_{k=1}^{3}\Big(\D{[\bBe]_{ij}}{x_k}v_k - \D{v_i}{x_k} \,[\bBe]_{kj} - [\bBe]_{ik} 
\,\D{v_j}{x_k}\Big).
\end{equation}
}It is important to remark that this time-derivative comes out naturally from our mathematical 
setting, and it is a correct representation of the modelling dynamics. Indeed, from Eq.\eqref{eq:Be_codeform} we gather 
$\bBe^{\uctd} = \bF \dot{\bH} \bF^T$, thus
$\bBe^{\uctd}$ vanishes if and only if no material 
remodelling occurs ($\dot{\bH}=0$ implies elastic, time-reversible, deformations).

When the free energy \eqref{eq:DensitaEnergiaPsiQ} is used, we find (see Appendix \ref{app:calculations} for more 
	details on the ordering tensor equation) the following explicit expressions for the pressure, the Cauchy 
	stress 
		tensor and the molecular field
	\begin{align}
	p(\rho) & = \rho^2 \D{\sio}{\rho} \label{eq:pressureQ} \\
	\bT & = -p(\rho) \,\bI + \rho \mu \big(\er^{-b\bQ}\bBe - \bI \big) - \rho k (\nabla \bQ)\odot(\nabla\bQ),
	\label{eq:TstressQ}\\
	\bh & = \rho \left(\D{\sigma_{\text{LdG}}}{\bQ} \right) - \divr\big(\rho k \nabla\bQ\big) 
	-\frac{1}{2}\rho \mu b \int_{0}^{1} \er^{\alpha\,b \bQ} (\er^{-b\bQ} \bBe) \er^{-\alpha\,b \bQ} d\alpha.
	\label{eq:moleculafieldQ}
	\end{align}

There are two types of governing equations. The first set of equations comprises balance laws that do not imply dissipation of energy. These, and the corresponding boundary conditions, are derived from the vanishing of the first four integrals in \eqref{eq:dissipazioneIntegrale_text}, for any test field. \comm{More precisely, we assume the local conservation of mass \eqref{eq:9a} and derive the remaining balance equations from \eqref{eq:dissipazioneIntegrale_text}, so that the equations for the density $\rho$, the velocity field $\bv$, and the ordering tensor $\bQ$ are}
\begin{subequations}
\begin{align}
\dot{\rho} + \rho\divr\bv & = 0, \label{eq:9a}\\
\rho \dot{\bv} & = \bb + \divr\bT, \label{eq:9b}\\
\str{\bh} & = 0, \label{eq:9c}
\end{align}
\label{eq:balanceeqns}
\end{subequations}
where $\str{\bh}= (\bh + \bh^T)/2 - \frac{1}{3}\tr(\bh) \bI$ is the traceless symmetric part of $\bh$.

 On the portion of the boundary where the traction is specified we have $\bT 
\bnu = \bt_{(\bnu)}$, where $\bnu$ is the outer unit normal. When no magnetic couple stress acts on the boundary, we 
read from \eqref{eq:dissipazioneIntegrale_text} the natural boundary condition for $\bQ$: 
\begin{equation}
\str{\left(\D{\sigma}{\nabla \bQ}\bnu \right)}=0.
\end{equation}

The second type of equations is associated with irreversible processes and follow from linear irreversible 
thermodynamics principles. These equations describe how the effective strain tensor $\bBe$ relaxes in time given the 
two competing effects: natural (viscous-like) relaxation towards the natural state dictated by $\bL$ and the 
external remodelling force given by the activity tensor $\bTa$. The simplification of Eq.\eqref{eq:evolution_general} 
leads to 
the following remodelling equation	
\begin{equation}
\tau\bBe^{\uctd} - \bBe^{-1} = \bTa - \er^{-b\bQ}\, 
\label{eq:evoluzione_Q}
\end{equation}
with $\tau$ a characteristic relaxation time. Eq.\eqref{eq:evoluzione_Q} guarantees that the dissipation 
\eqref{eq:dissipazioneIntegrale_text} is always greater or equal to zero, so that the second principle of 
thermodynamics is satisfied.

When $\tau \gg 1$, i.e., for very long relaxation times, the material response is elastic, there is no dissipation of 
energy and, to leading order, \eqref{eq:evoluzione_Q} becomes $\bBe^{\uctd}=0$. This is the condition to impose if we 
want to reproduce an elastic, non-dissipative, response in our theory. By contrast, when $\tau \ll 1$, i.e., for very 
short relaxation times compared to the characteristic times of the macroscopic motion (measured by $1/\|\nabla\bv\|$), 
the tensor $\bBe$ quickly relaxes to a new stationary state. In such a case, we reproduce ideal and viscous-flow 
behaviours. This approximation will be explored in more detail in \S\ref{sec:fast_relaxation}. In intermediate regimes 
we recover a viscoelastic material response.

Unlike classical theoretical models of plasticity, the evolution equation \eqref{eq:evoluzione_Q} is written in 
terms of the spatial (Eulerian) tensor field $\bBe$. In so doing, we regard $\bBe$ as a state variable that 
characterize the stress state of the material (via Eq.\eqref{eq:TstressQ}) and eliminate from the theory the 
arbitrariness due to the choices of reference and natural configuration \cite{13Volokh,19Rubin,Rubin}.

	% \begin{align}
	% \D{}{\bQ}\,\tr\left(\bPsi^{-1}\bBe\right) 
	% =  - b \int_{0}^{1} \er^{(\alpha-1)\,b \bQ}\,\bBe \,\er^{-\alpha\,b \bQ} d\alpha
	% \end{align}
	% \begin{equation}
	% \D{}{\bQ}\,\tr\left(\bPsi^{-1}\bBe\right) =  - \left(\D{\bPsi}{\bQ} \right): \left[ 
	%\bPsi^{-1}-\bbD\left(\bPsi^{\uctd}\right)\right]
	% \end{equation}
	
\section{Active fluid approximation}
\label{sec:fast_relaxation}

When the relaxation time $\tau$ is very short compared to the characteristic times of deformation, measured by 
$1/\|\nabla \bv\|$, the material effectively behaves as an ordinary fluid since the material reorganization is much 
faster than the deformation. Hence, our theory reduces to that of an active liquid crystal. For a simpler comparison 
with existing theories, we also assume low activity. Specifically, we assume that $\bBe$ is just a small correction of its equilibrium value in the passive case
\begin{equation}
\bBe = \bL + \bB_1, \qquad \text{ with } \qquad \|\bB_1\| = O(\eps), \qquad \|\bTa\| = O(\eps),
\label{eq:fast_relaxation_assumptions}
\end{equation}
where $\eps=\tau\|\nabla\bv\|$, and we recall that $\bL=\er^{b\bQ}$. The substitution of 
Eq.\eqref{eq:fast_relaxation_assumptions} in 
Eq.\eqref{eq:evoluzione_Q}, yields, to first order,
\begin{equation}
\bL^{-1}\bBe = \bI + (\bTa - \tau \bL^{\uctd})\bL + O(\eps^2)\, .
\label{eq:LB_fast}
\end{equation}
Eq.\eqref{eq:LB_fast} can be inserted in Eq.\eqref{eq:TstressQ} so that the stress tensor for the active fluid reads
\begin{equation}
\bT = -p(\rho) \,\bI + \rho \mu  (\bTa - \tau \bL^{\uctd})\bL - \rho k (\nabla \bQ)\odot(\nabla\bQ),
\label{eq:Tstress2}
\end{equation}
which actually shows an active term, $\rho \mu \bTa \bL$, that can be chosen to share the nematic symmetry with 
$\bQ$, in agreement with the classical active liquid crystal theories. The passive term, $-\rho \mu 
\tau\bL^{\uctd}\bL$, generates the nematic liquid crystal viscous stress and allows us to identify the Leslie viscosity 
coefficients in terms of the material parameters $\rho$, $\mu$, $a$, and 
$\tau$ \cite{16Turzi}, and their dependence on the degree of order $S$. It is worth noticing that no previous knowledge 
of the viscous terms compatible with nematic symmetry is required to construct the stress tensor. The correct 
dependence on the director and its derivatives appear naturally in the expansion as a consequence of the calculations.

Likewise, no integrity bases of scalar invariants constructed from $\bQ$ and its derivatives is needed to find the 
evolution equation for the ordering tensor $\bQ$. This is particularly interesting since, in a traditional approach, 
there is in principle no limitation on the number of times $\bQ$ can appear in the ordering tensor equation 
\cite{12Virga_book}. By contrast, in our case, the only terms appearing in this equation are those coming from the 
derivatives of the free energy density.

For all practical purposes, it is probably more convenient to use the integral formula \eqref{eq:moleculafieldQ} to calculate the molecular field. However, in order to see what combinations of $\bQ$ (and its derivatives) appear in the ordering tensor equation, and for a more transparent comparison with the existing theories, it is helpful to recast $\bL^{\uctd}$ in terms of $\bQ$ and its derivatives. This is possible if we further assume that $b \ll 1$, i.e., there is a weak coupling between orientational order and natural strain. In such a case we have $\bL = \er^{b\bQ} \approx \bI + b \bQ$ which still yields $\det\bL = 1$ to first order in $b$. Hence, we recover the classical linear functional dependence of $\bL$ on $\bQ$ (see the discussion on p.59--61 of \cite{Warner}).

The integral in the molecular field Eq.\eqref{eq:moleculafieldQ} can be calculated explicitly under this assumption, as reported in Appendix \ref{app:calculations_fast}. The ordering tensor equation \eqref{eq:9c} now reads
\begin{align}
%& \rho \left( A \bQ - B \Big(\bQ^2 - \frac{1}{3}\tr(\bQ^2)\,\, \bI \Big) + C\,\tr(\bQ^2) \bQ \right) \notag 
& \rho \str{\left(\D{\sigma_{\text{LdG}}}{\bQ} \right)} \notag 
- \divr\left(\rho k  \nabla \bQ\right) \notag \\
& - \frac{1}{2}\rho \mu b
\str{\Big( \bTa +2\tau \bD +\frac{b}{2}\big(\bQ\bTa + \bTa\bQ\big) 
+ 2b\tau \big(\bQ\bD + \bD \bQ \big)
- b\tau \bQd \Big)} = 0,
\label{eq:Q_fast}
\end{align}
so that we recover the same combination of terms ($\bD$, $\bQ\bD + \bD \bQ$, and $\bQd$) that is traditionally obtained when the dissipation function is truncated at second degree in $\bQ$, $\bQd$ \cite{94Beris,98Qian,17Yeomans} (see \cite{04Virga,12Virga_book} for a fuller discussion). It is worth noticing that the activity remodelling tensor $\bTa$ is explicitly included also in the ordering tensor equation \eqref{eq:Q_fast}.

\subsection{Active spontaneous flow in the isotropic phase}
\label{sec:applications}
\begin{figure}[!h]
	\begin{center}
		\includegraphics[width=0.4\linewidth]{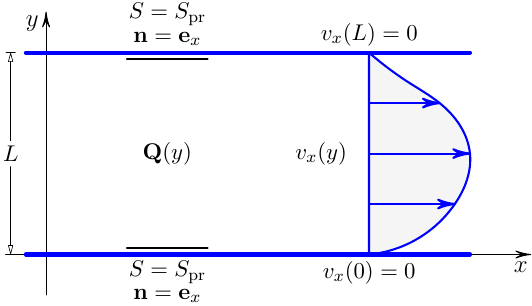}
	\end{center}
	\caption{Schematic representation of the channel geometry. Translational invariance along the $x$-axis ensures that all the unknown fields only depend on the transverse variable $y$, and that $v_y=0$. We assume no-slip walls at $y=0$, $y=L$, and planar boundary conditions for the nematic director. We set $S=\spr$ at the boundary.}
	\label{fig:spflow}
\end{figure}

A distinguishing feature of active nematics is that they have a natural tendency to flow when the activity 
parameter overcome a typical critical value \cite{05voituriez,09edwards,11Giomi,12Giomi,16Turzi,17Shendruk}. It is then 
natural to test our model against such a prediction and, in particular, to investigate the interplay between activity 
and preferred order. To this end, we study the spontaneous flow of an active nematic fluid in a simple channel 
geometry. We consider here only high-dense suspensions, where density variations are negligible and the material is 
nearly incompressible. Therefore we neglect the important effect of particle concentration at the boundaries that is 
observed for low-dense active suspensions \cite{14Fily}.
For simplicity, we consider a system with translational invariance along $x$, and we assume that the unknown fields, 
$v_x$, $Q_{xx}$ and $Q_{xy}$, depend only on the transverse variable $y$ (see Fig. \ref{fig:spflow}). We recall that in 
our two-dimensional example $\bQ$ and $\str{\bh}$ are $2 \times 2$ symmetric and traceless matrices, so that 
$Q_{yx}=Q_{xy}$, $Q_{yy}=-Q_{xx}$, $\str{h}_{yx}=\str{h}_{xy}$ and $\str{h}_{yy}=-\str{h}_{xx}$. The degree of 
order, $S$, and the director, $\bn$, are then derived from the eigenvalues and eigenvectors of $\bQ$.

The nematic degree of order in the active system is governed by a Landau-de Gennes potential, whose minimum is denoted 
with $\spr$, the preferred value of the degree of order $S$. In our thin channel geometry, we posit
\begin{equation}
\sigma_{\text{LdG}}(\bQ) 
= \tfrac{1}{2}A\tr(\bQ^2) + \tfrac{1}{4}C\tr(\bQ^2)
= \frac{k}{\xi^2} S^2 (S^2 -2\spr),
\label{eq:LdG2D}
\end{equation}
where we have parametrized $\sigma_{\text{LdG}}(\bQ) $ using the nematic coherence length $\xi = \sqrt{4k/C}$, and the 
preferred degree of order is $\spr=\sqrt{-2A/C}$. The nematic coherence length compares the strength of the elastic and 
thermodynamic contributions to the free-energy. It characterizes the size of the domains, where the degree of 
orientation may be different from the preferred value $\spr$. We have omitted the cubic term in Eq.\eqref{eq:LdG2D} 
since we are 
using the two-dimensional ordering tensor $\bQ$, which has only one quadratic invariant. In the nematic phase, the 
Landau-de Gennes potential has two symmetric minima in $S= + \spr$ and $S= - \spr$, and a maximum in $S=0$. The 
isotropic phase corresponds to $S=0$. To see the effect of preferred order on the flow, we perform the linear analysis 
using a generic value for $\spr$. The isotropic case will be discussed by setting $\spr=0$ in the final formulas.

The equations of motions are given as in \eqref{eq:9b} and \eqref{eq:9c}, where the Cauchy stress tensor and the 
molecular field are given as in the active fluid approximation described in \S\ref{sec:fast_relaxation}. It is worth 
mentioning that \eqref{eq:9a} is automatically satisfied due to incompressibility. Furthermore, due to stationarity and 
translational invariance along the channel, Eqs.\eqref{eq:9b} and \eqref{eq:9c} simply read $\D{T_{xy}}{y} = 0$, 
$\D{T_{yy}}{y} = 0$ and $\str{h_{xy}}=0$, $\str{h_{yy}}=0$. We also assume no-slip boundary conditions for the velocity 
field 
$v_x(y)$, tangential conditions for the director and $S(0)=\spr$, $S(L)=\spr$ at the channel walls.

The momentum balance Eq.\eqref{eq:9b}, in the stationary case and with $\bT$ as given in \eqref{eq:Tstress2}, yields
\begin{subequations}
\begin{align}
4 b \tau  Q'_{xx} v_x'+b \zeta  Q'_{xy}+2 \tau  \left(2 b Q_{xx}-1\right) v_x'' & = 0 \label{eq:20a} \\
\rho  Q'_{xy} \left(b \mu  \tau  v_x'-4 k Q''_{xy}\right)+\frac{1}{2} \rho  Q'_{xx} \left(b \zeta  \mu -8 k Q''_{xx}\right)+b \mu  \rho  \tau  Q_{xy} v_x''- p' & = 0 \label{eq:20b}
\end{align}
\label{eq:20}
\end{subequations}
In agreement with the active fluid approximation of \S\ref{sec:fast_relaxation}, we have used $\bL\approx \bI + b \bQ$ and retained only terms up to first order in $b$. Translational invariance implies that the unknown fields, namely $v_x$, $Q_{xx}$, and $Q_{xy}$, are functions only of the transverse variable $y$. We have chosen the simplest form for the activity tensor, i.e., $\bTa = -\tfrac{1}{2}\zeta \bI$. Eq.\eqref{eq:20b} can be used to find the pressure function $p(y)$, but it will not be used in the rest of the paper. 

Two more scalar equations are obtained from the order tensor equation Eq.\eqref{eq:9c} (or, more directly, 
Eq.\eqref{eq:Q_fast}), which, to order $O(b)$, reads
\begin{subequations}
\begin{align}
k Q''_{xx}+\frac{2 k Q_{xx} \left(\spr^2-4 Q_{xy}^2\right)}{\xi ^2}-\frac{8 k Q_{xx}^3}{\xi ^2} & = 0 , \label{eq:21a}    \\
k Q''_{xy}+\frac{2 k Q_{xy} \left(\spr^2-4 Q_{xx}^2\right)}{\xi ^2}-\frac{8 k Q_{xy}^3}{\xi ^2} + \frac{1}{2} b \mu  
\tau  v_x' & =0. \label{eq:21b}
\end{align}
\label{eq:21}
\end{subequations}

Eqs.\eqref{eq:20a},\eqref{eq:21a}, and \eqref{eq:21b} are three nonlinear equations in the unknowns $v_x$, $Q_{xx}$ and 
$Q_{xy}$ that are in general difficult to solve. However, it is straightforward to check that $Q_{xx} = \spr/2$, 
$Q_{xy} = 0$, $v_x = 0$ is always a trivial solution, for any value of the parameters. Furthermore, above a critical 
threshold for the activity parameter $\zeta$, a bifurcation occurs, and the trivial solution is no longer unique. 

The critical condition is found by performing a linear stability analysis about the trivial solution. To find the 
linear equations, we perturb the trivial solution and set $Q_{xx} = \spr/2 + \delta Q_{xx}$, $Q_{xy} = \delta Q_{xy}$, 
$v_x = \delta v_x$. To first order, Eqs.\eqref{eq:20a}, \eqref{eq:21a}, and \eqref{eq:21b} yield
\begin{subequations}
\begin{align}
b \zeta  \,\delta Q_ {xy}' + 2 \tau  (b \spr-1) \,\delta v_x'' & = 0 ,\label{eq:22a}    \\
\delta Q_{xx}''-\frac{4 \spr^2 \,\delta Q_{xx}}{\xi ^2} & = 0, \label{eq:22b} \\
k \,\delta Q_{xy}'' + \frac{1}{2} b \mu  \tau  \,\delta v_x' & = 0. \label{eq:22c}
\end{align}
\label{eq:22}
\end{subequations}
Eq.\eqref{eq:22b} is decoupled from the other two equations and can be solved immediately: the only solution that 
satisfies both boundary conditions is $\delta Q_{xx}=0$. By contrast, it is easy to show that Eqs.\eqref{eq:22a}, 
\eqref{eq:22c} admit real non-trivial solutions whenever $\zeta/(1-b \spr) > 0$ and the following critical condition is 
satisfied
\begin{equation}
\sqrt{\frac{\zeta  \mu }{4k}} \, \frac{bL}{\sqrt{1-b \spr}} = 2 \pi n,
\label{eq:critical}
\end{equation}
where $n$ is a non-vanishing integer. The condition $\zeta/(1-b \spr) > 0$ implies that we can investigate spontaneous 
flow even in the absence of a Landau-de Gennes potential, i.e., when $\spr=0$, provided that $\zeta > 0$. 

Furthermore, for every $n$, we find that the null space of the linear operator is 
two-dimensional, so that there are two possible modes of bifurcations in agreement with the 
linear stability analysis given in \cite{16Turzi} and the numerical studies of 
Refs.\cite{07MarenduzzoPRL,07MarenduzzoPRE}. 

\begin{align}
\delta Q_{xy} & = 2 \pi n \alpha \sin \left(\frac{2 \pi  n y}{L}\right) 
- \frac{16 \pi ^2 n^2 (1-b \spr)}{b \zeta } \beta\sin ^2\left(\frac{\pi  n y}{L}\right)
\label{eq:bifmodeQ}\\
\delta v_x & = \frac{L}{\tau} \left[ \frac{b \zeta}{1 -b \spr} \alpha \sin ^2\left(\frac{\pi  n y}{L}\right)
+ 2 n\pi \beta \sin \left(\frac{2 \pi  n y}{L}\right)\right], \label{eq:bifmodev}
\end{align}
where $\alpha$ and $\beta$ are the two arbitrary mode amplitudes that cannot be determined from this simple linear 
analysis. Using Eqs.\eqref{eq:bifmodeQ},\eqref{eq:bifmodev} it is possible to derive the qualitative behaviour of $S$, 
$v_x$ and the angle $\theta$ as a function of $y$ (for an approximate quantitative analysis we need to know the 
amplitudes $\alpha$, $\beta$).

By contrast, other works \cite{05voituriez,09edwards,12Giomi,15Yeomans,20Yeomans} only predict a single bifurcation 
mode, i.e., a sinusoidal modulation with a node at the centre of the channel. The tumbling parameter does not play a 
significant role in our theory, while a key parameter is the coupling constant $b$. This is again a difference with 
respect to the models inspired by Beris-Edwards theory of liquid crystals (see for instance 
\cite{14Giomi,15Yeomans,20Yeomans,21Vitelli}).

\begin{figure}[!h]
\begin{center}
	\includegraphics[width=0.6\linewidth]{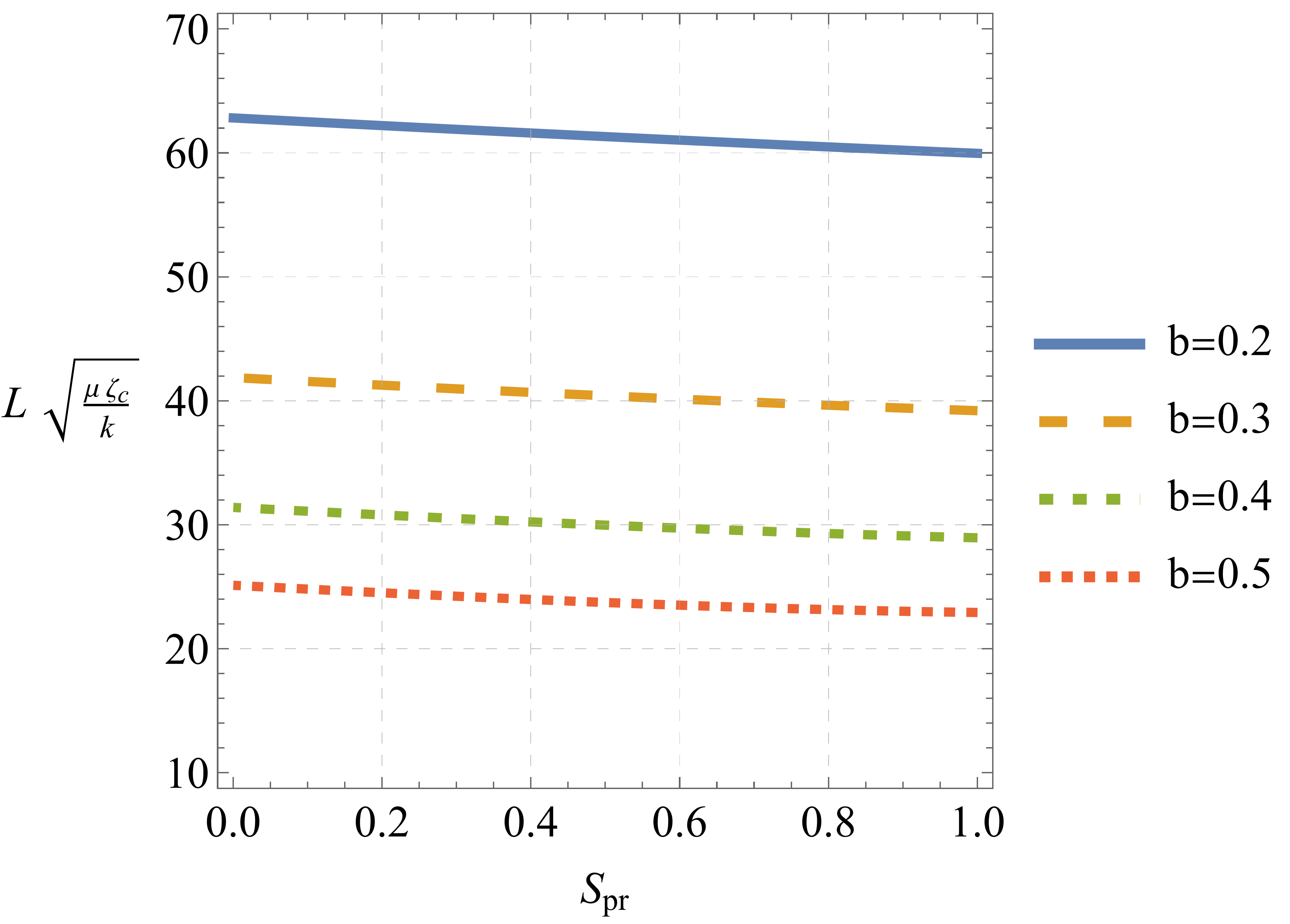}
\end{center}
\caption{Critical activity, $\zeta_c$, as a function of the preferred degree of order, $\spr$, as given in 
Eq.\eqref{eq:critical} with $n=1$, for different values of the coupling parameter $b$. There is a finite $\zeta_c$ even 
when $\spr=0$, so that the preferred phase is isotropic.}
\label{fig:critical activity}
\end{figure}

For extensile materials, with $\zeta >0$, it is possible to find a critical activity also when the preferred nematogens 
alignment is isotropic ($\spr=0$). In order to analyse this case, we plot in Fig. \ref{fig:critical activity} the 
critical activity, $\zeta_c$ as given in \eqref{eq:critical} with $n=1$, as a function of the preferred degree of 
orientation, $\spr$, for different values of the coupling parameter $b$. As expected, $\zeta_c$ increases 
with decreasing $b$ and diverges when $b\to 0$. In physical terms this means that if there is no coupling between 
orientational order and natural strain then no spontaneous flow arises. However, it is interesting to observe that even 
in the absence of any preferred orientation, when $\spr=0$, and the material is naturally isotropic, the critical 
activity is still finite, so that it is possible to observe an activity-induced spontaneous flow even when 
$\sigma_{\text{LdG}}=0$. This is in agreement with some recent results \cite{15Yeomans,20YeomansB,21Giomi}, which study 
the instability in Beris-Edwards theory of active nematic liquid crystals or flocking dynamics with Toner-Tu theory.

\section{Activity-induced actuation}
\label{sec:elastic}
\comm{
Our model can in principle describe a range of viscoelastic behaviours that are present in a number of polymeric and biological materials including entangled protein filaments and biological and artificial muscles. To encompass this diversity, the relaxation time, $\tau$, may vary from $\tau \ll 1$ where the polymeric dynamics are rapid and only contribute to the viscosity of the fluid, to $\tau \to +\infty$, where the
polymer effectively acts as an elastomeric solid. 

In \S\ref{sec:fast_relaxation} we discussed the first approximation. Here, we briefly explore one simple consequence of the active elastic approximation, namely, the spontaneous stretch induced by an active term. In general, it is more difficult to study the elastic limit. In fact, $\bBe^{\uctd}$ becomes vanishingly small while $\tau \to +\infty$, so that  $\tau \|\bBe^{\uctd}\|=O(1)$. This means that all the terms in \eqref{eq:evoluzione_Q} are of the same order and no dominant balance argument can be applied to simplify the equation. A possible simplification is to use a small displacement approximation and write 
\begin{equation}
\bBe = \bL + \bB_1.
\end{equation}
This is similar to the assumption \eqref{eq:fast_relaxation_assumptions} in \S\ref{sec:fast_relaxation}, however this is now not due to a fast remodelling that makes $\bBe$ always close to the natural configuration $\bL$, but rather to a small displacement approximation, which may be due to a small active contribution. This may not be valid at large times since soft matter materials usually undergo large deformations and $\bBe$ can in principle be large in the elastic regime (despite $\bBe^{\uctd}$ being small).

We consider a homogeneous infinite material (i.e., we are not concerned with boundary conditions) at rest and explore the effect of activity on its equilibrium configuration. Within the small displacement approximation, the zeroth-order remodelling equation \eqref{eq:evoluzione_Q} simply yields $L^{\uctd}=0$ which is automatically satisfied for homogeneous equilibrium states (i.e., with constant $\bQ$, and $\bv=0$).
%To this end,  In the small displacement approximation $\bBe$ is just a small perturbation of $\bL$,
To find the contribution of activity we must proceed to the next order of approximation, which gives
\begin{equation}
\tau \bB_1^{\uctd} + \bL^{-1}\bB_1\bL^{-1} = \bTa.
\label{eq:elastic_B1}
\end{equation}
Therefore, in the elastic case, an active equilibrium configuration with $\bv=0$ is possible only if $\bB_1 = \bL\bTa \bL$. We need to check whether this solution is compatible with the balance of linear momentum, $\divr \bT = 0$ and the director equation. To first order, the Cauchy stress tensor \eqref{eq:TstressQ} becomes
\begin{equation}
\bT = -p(\rho) \bI + \rho\mu \, \bL^{-1}\bB_1 = -p(\rho) \bI + \rho\mu \, \bTa \bL,
\end{equation}
where, as usual in nematic elastomers, we have neglected the contribution of the Ericksen stress tensor (which, anyway, vanishes in a homogeneous state). We also take $S=$constant and write $\bn = (\cos\theta(y),\sin\theta(y),0)$. For concreteness, we choose the active tensor along a specific axis: $\bTa  =-\frac{1}{2}\zeta \be_x \tp \be_x$. After some algebra, the equilibrium equation $\divr \bT = 0$ and the director equation require
\begin{align}
\frac{a^3-1}{2a} \zeta  \mu  \rho  \theta '(y) \cos (2 \theta (y)) = 0, \qquad 
p'(y) = 0, \qquad  
\sin (2 \theta (y)) = 0.
\end{align}
Hence, a homogenous stationary solution is possible with either $\theta=0$ or $\theta=\pi/2$, i.e., $\bn=\be_x$ or $\bn \perp \be_x$. In the first case we have the natural deformation 
\begin{equation}
\bBe(\bn=\be_x) = a^2(1-a^2 \zeta/2)\,\be_x \tp \be_x + \frac{1}{a}(\bI-\be_x \tp \be_x),
\end{equation}
while in the second 
\begin{equation}
\bBe(\bn=\be_y) = \left(\frac{1}{a}-\frac{\zeta }{2 a^2}\right)\,\be_x \tp \be_x + a^2 (\be_y \tp \be_y) 
+ \frac{1}{a}(\be_z \tp \be_z) .
\end{equation}
In the presence of activity, both the director and the elastic matrix evolve in agreement with the activity tensor. There are homogeneous equilibrium solutions, which have residual stress and are stretched or contracted, with respect to the natural state $\bBe=\bL$, according to the action of the activity tensor. 
}

\section{Conclusions}
\label{sec:conclusions}

Many continuum theories seem to effectively describe out of equilibrium dynamics of active nematics in terms of a 
large number of phenomenological parameters. Furthermore, such parameters are difficult to determine from microscopic 
information or experimental data, in particular in active systems where energy is injected at the microscopic units.
Most active nematic responses depend on the competition between active stresses that promote director or velocity 
gradients and viscoelastic stresses that resist them. As a consequence of this interplay, experimental measurements 
often access only non-trivial combinations of hydrodynamic parameters, and it is very difficult to assess the importance 
of each single parameter. Even direct measurements that rely on controlled flow experiments are difficult to devise if 
the underlying flows are chaotic. As a result, there is still no consensus on which macroscopic description is the most 
physically sound.

In this paper we have presented a minimal continuum theory with few, but not 
fewer, key parameters having a clear physical interpretation. The model we propose is based on a kinematic coupling between nematic elastomer shape-tensor and de Gennes ordering tensor, and material remodelling dynamics. Hence, it is able to account for viscoelastic 
effects, partial degree of nematic order and defect dynamics. In order to explore the first implications of the theory, we studied the dynamical properties of thin films of active 
nematic fluids confined in a channel geometry. Above a critical activity, a rich variety of complex behaviours is 
observed. In particular,  it is possible to observe spontaneous flow of particles even in the 
absence of any thermodynamic potential that could favour the alignment of the rodlike units. Therefore, activity 
itself can give rise to local flows which in turn gives rise to nematic order, and this enhances the effect of activity 
and the flows itself. This effect is important to explain the origin of nematic order in active systems and hence the 
physics of biological matter. In fact, most theoretical models assume the presence of nematic order by prescribing a 
Landau-de Gennes potential, but it is not clear what could be the microscopic origin of this potential, especially 
since many active systems do not maintain nematic order when the activity magnitude tends to zero.

%\funding{This research was partially supported by GNFM of Italian Istituto Nazionale di Alta Matematica.}

\section*{acknowledgments}
This research was partially supported by GNFM of Italian Istituto Nazionale di Alta Matematica.

%\disclaimer{Insert disclaimer text here.}

\appendix
\renewcommand{\theequation}{\thesection\ \arabic{equation}}
\section{Derivation of the governing equations}
\label{app:equations}
\comm{
Let us first state two lemmas that will be used to simplify the calculations and which we state without proof.
\begin{lemma} Let $f$ be a function depending only on $\bFe$ (or $\bBe$). Then,
\begin{enumerate}
\item[(1)] $\displaystyle\D{f}{\bF}\bF^{T} = \D{f}{\bFe}\bFe^{T} = 2\D{f}{\bBe}\bBe$
\item[(2)] $\displaystyle\bF^{-T}\D{f}{\bH}\bF^{-1} = \D{f}{\bBe}$
\end{enumerate}
\label{teo:lemma1}
\end{lemma}
\begin{lemma} Let $f=f(\rho)$ be a function of $\rho$ only. Then, $\displaystyle\D{f}{\bF}\bF^{T} = -\rho\D{f(\rho)}{\rho} \bI$.
\label{teo:lemma2}
\end{lemma}
}
Let us define the expended power by the external forces, the kinetic energy and the free energy as
\begin{align}
\Wext & := \int_{\Pt} \bb\cdot \bv \,\,dv + \int_{\partial\Pt} \bt_{(\bnu)}\cdot \bv \,\,da \notag \\
& + \int_{\Pt} \bE \cdot \dot{\bQ} \,\,dv + \int_{\partial\Pt} \bE_{(\bnu)} \cdot \dot{\bQ} \,\,da \notag \\
& + \int_{\Pt} \bTa\cdot \bBe^{\uctd} \,\,dv, \label{eq:Wext} \\
K + \FF & := \int_{\Pt} \left(\frac{1}{2}\rho \bv^2 + \rho \si(\rho,\bBe,\bQ,\nabla\bQ) \right) \,dv,
\end{align}
where $\bv$ is the velocity field,\comm{$\Pt$ is a spatial region that convects with the body}, and 
\begin{equation}
\bBe^{\uctd} := (\bBe)\Ldot{\phantom{I}} - (\nabla\bv) \,\bBe - \bBe \,(\nabla\bv)^T = \bF\Ldot{\bH}\bF^{T},
\label{eq:Be_codeform}
\end{equation}
is the codeformational derivative\footnote{Also known as upper-convected time derivative, upper-convected rate or 
contravariant rate.} \cite{Joseph,Gurtin}, a frame-indifferent time-derivative of $\bBe$ relative to a convected 
coordinate system that moves and deforms with the flowing body. The unit vector $\bnu$ is the external unit normal to 
the boundary $\partial\Pt$; $\bb$ is the external body force,  $\bt_{(\bnu)}$ is the external traction on the bounding 
surface $\partial\Pt$. The tensor fields $\bE$ and $\bE_{(\bnu)}$ are the generalized body and contact force densities 
conjugate to the microstructure: they have to be included for example in the presence of an external magnetic field, 
but for simplicity they will be set to zero in the following derivations. The last term in Eq.\eqref{eq:Wext} 
represents the power expended by the \emph{remodelling generalised force} $\bTa$ \cite{17tur}. It must be noted that 
$\bTa$ is the only term 
conjugate to the remodelling velocity field $\bBe^{\uctd}$. By contrast, the classical active stress $\sa = -\zeta\bQ$ is 
paired with the macroscopic velocity gradient $\nabla\bv$. It is clear from the identity 
$\bBe^{\uctd}=\bF\Ldot{\bH}\bF^{T}$ that 
$\bBe^{\uctd}$ has 
the right properties to represent the kinematics of reorganization: (1) it is frame-invariant, (2) it vanishes whenever 
the deformation is purely elastic, i.e., $\bBe^{\uctd} = 0$ if and only if $\Ldot{\bH}=0$. The same derivative also 
appears in the three-dimensional models for Maxwell viscoelastic fluids \cite{Joseph}. 

To calculate the dissipation we first calculate the material time-derivative of $K+\FF$
\begin{align}
\dot{K} + \dot{\FF} 
&= \int_{\Pt} \left(\rho\dot{\bv}\cdot \bv 
+ \rho\D{\si}{\bF}\cdot \Ldot{\bF} 
+ \rho\D{\si}{\bH}\cdot \Ldot{\bH}
+ \rho \D{\si}{\bQ}\cdot \Ldot{\bQ} 
+ \rho \D{\si}{\nabla\bQ}\cdot (\nabla\bQ)\dot{\phantom{i}} \right) dv \notag \\
&= \int_{\Pt} \left(\rho\dot{\bv}\cdot \bv 
+ \rho\D{\si}{\bF}\bF^T\cdot \nabla\bv 
+ \rho\D{\si}{\bBe}\cdot \bBe^{\uctd}
+ \rho \D{\si}{\bQ}\cdot \Ldot{\bQ} 
+ \rho \D{\si}{\nabla\bQ}\cdot (\nabla\bQ)\dot{\phantom{i}} \right) dv,
\end{align}
\comm{where, to simplify the calculations, we have implicitly used the local conservation of mass $\dot{\rho} + \rho\divr\bv = 0$.}
The last term in the integral is simplified by using the following identities (with $\bF_{jk}=\partial x_j/\partial 
X_k$)
\begin{align}
[(\nabla\bQ)\dot{\phantom{i}}]_{hi,j} & 
= \frac{d}{dt} \left(\D{Q_{hi}}{X_k}\D{X_k}{x_j}\right) 
= (\nabla \Ldot{\bQ})_{hi,j} + \D{Q_{hi}}{X_k}\left(-\bF^{-1}\nabla \bv \right)_{kj}
= [\nabla \Ldot{\bQ} - (\nabla \bQ)(\nabla \bv)]_{hi,j}, \\
\rho\D{\si}{\nabla\bQ}\cdot (\nabla\bQ)\dot{\phantom{i}} 
& = \rho\D{\si}{\nabla\bQ}\cdot (\nabla\Ldot{\bQ}) - \rho\D{\si}{\nabla\bQ}\cdot (\nabla \bQ)(\nabla \bv) \notag \\
& = \rho\D{\si}{Q_{hi,j}}\,\Ldot{Q}_{hi,j} - \rho\D{\si}{Q_{hi,j}}\, 
Q_{hi,k}\,v_{k,j} \notag \\
& = \left(\rho\D{\si}{Q_{hi,j}}\,\Ldot{Q}_{hi} \right)_{,j} 
- \left(\rho\D{\si}{Q_{hi,j}} \right)_{,j} \,\Ldot{Q}_{hi}
- \rho Q_{hi,k}\,\D{\si}{Q_{hi,j}}\,v_{k,j} \notag \\
& = \divr\left(\rho \Ldot{\bQ} \odot \D{\si}{\nabla\bQ}\right) 
- \divr\Big(\rho\D{\si}{\nabla\bQ}\Big) \cdot \Ldot{\bQ}
- \left(\rho(\nabla \bQ)\odot\D{\si}{\nabla\bQ}\right)\cdot \nabla \bv,
\end{align}
where, following \cite{04Virga}, the circled dot means the contraction of the first two indexes, so for instance 
$(\nabla\bQ \odot \nabla\bQ)_{jk} 
= Q_{hi,j} Q_{hi,k}$. Hence, we obtain
\begin{align}
\dot{K} + \dot{\FF} 
&= \int_{\Pt} \left(\rho\dot{\bv}\cdot \bv 
+ \bT\cdot \nabla\bv 
+ \rho\D{\si}{\bBe}\cdot \bBe^{\uctd}
+ \bh \cdot \Ldot{\bQ} \right) dv
+ \int_{\partial\Pt} \left(\rho \D{\si}{\nabla \bQ}\bnu \right) \cdot \Ldot{\bQ}\, da,
\end{align}
where we have defined the Cauchy stress tensor and the molecular field as
\begin{align}
\bT & = \rho\D{\si}{\bF}\bF^{T} - \rho(\nabla \bQ) \odot \D{\si}{\nabla\bQ} 
= -\rho^2 \D{\si}{\rho} \bI + \rho\D{\si}{\bFe}\bFe^{T} - \rho(\nabla \bQ)\odot\D{\si}{\nabla\bQ}, \label{eq:defT}\\
\bh & = \rho\D{\si}{\bQ} - \divr\Big(\rho\D{\si}{\nabla\bQ}\Big). \label{eq:defh}
\end{align}
A further simplification using divergence theorem yields the final expression for the dissipation
\begin{align}
\mathcal{D} & = \Wext - \dot{K} - \dot{\FF} 
= \int_{\Pt} \left(\bb - \rho \dot{\bv} + \divr\bT \right) \cdot \bv \,\,dv \notag \\
& + \int_{\partial\Pt} \left(\bt_{(\bnu)} - \bT\bnu \right) \cdot \bv \,\,da 
- \int_{\Pt} \bh \cdot \Ldot{\bQ} \,\,dv \notag \\
& - \int_{\partial\Pt} \left(\rho \D{\si}{\nabla \bQ}\bnu \right) \cdot \Ldot{\bQ}\, da
+ \int_{\Pt}\left(\bTa-\rho\D{\si}{\bBe}\right)\cdot \bBe^{\uctd}\,\, dv,
\label{eq:dissipazioneIntegrale}
\end{align}
where we have used $\bE=0$ and $\bE_{(\bnu)}=0$. The key assumption of our derivation is that only the 
last integral in \eqref{eq:dissipazioneIntegrale} can be responsible of a positive dissipation. In physical terms, 
positive dissipation is uniquely associated with material remodelling. Hence, balance equations \eqref{eq:balanceeqns} 
(and corresponding boundary terms) are obtained by setting to zero the 
terms conjugate to the arbitrary kinematic fields $\bv$ and $\Ldot{\bQ}$. Furthermore, since $\Ldot{\bQ}$ is symmetric 
and 
traceless, $\bh \cdot \Ldot{\bQ} = \str{\bh} \cdot \Ldot{\bQ}$, where $\str{\bh}= (\bh + \bh^T)/2 - \frac{1}{3}\tr(\bh) 
\bI$ is the symmetric traceless part of $\bh$, so that it is sufficient to posit $\str{\bh}=0$ to derive the balance 
equation. 

After inserting Eqs.\eqref{eq:balanceeqns} in \eqref{eq:dissipazioneIntegrale}, the dissipation simplifies to
\begin{align}
\mathcal{D} = \int_{\Pt}\left(\bTa-\rho\D{\si}{\bBe}\right)\cdot \bBe^{\uctd}\,\, dv,
\label{eq:dissipazioneIntegrale2}
\end{align}
and its positivity is guaranteed if we take the evolution equation \eqref{eq:evoluzione_Q}. More generally, it is 
sufficient to take 
\begin{equation}
\bbD (\bBe^{\uctd}) = \bTa-\rho\D{\si}{\bBe}\, .
\label{eq:evolution_general}
\end{equation}
where $\bbD$ is a positive-definite fourth-rank tensor with the major symmetries, i.e., such that $\bA \cdot \bbD\bA 
>0$ for any symmetric $\bA \neq 0$. See \cite{16Turzi} for a more detailed discussion 
about the structure of the tensor $\bbD$ and the possible relaxation times. The evolution equation given in the text is 
the simplest 
possible choice with only one relaxation time, after the substitution of 
\begin{equation}
\rho\D{\si}{\bBe} = \frac{1}{2}\mu \left(\bPsi^{-1} - \bBe^{-1}\right) 
= \frac{1}{2}\mu \left(\er^{-b \bQ} - \bBe^{-1}\right).
\end{equation}

\section{Ordering tensor equation}
\label{app:calculations}

In order to find $\str{\bh}$ we use Eqs.\eqref{eq:LdG},\eqref{eq:Frank} to calculate the following Fréchet derivatives
\begin{align*}
\D{\sigma_{\text{LdG}}(\bQ)}{\bQ} & =  A \bQ - B \bQ^2 + C\,\tr(\bQ^2) \bQ , \\
\str{\left(\D{\sigma_{\text{LdG}}(\bQ)}{\bQ}\right)} & =  A \bQ - B \Big(\bQ^2 - \frac{1}{3}\tr(\bQ^2)\,\, 
\bI \Big) + C\,\tr(\bQ^2) \bQ , \\
\D{\sigma_{\text{el}}(\bQ,\nabla\bQ)}{\nabla\bQ} & = k(\rho) \nabla \bQ .
\end{align*}
More involved is the calculation of the derivative of the neo-Hookean term $g(\bQ)=\tfrac{1}{2}\mu (\er^{-b \bQ}\cdot 
\bBe)$. We will use Feynman's formula for the derivative of the matrix exponential (Eq.(6), \cite{51Feynman} or 
\cite{67Wilcox})
\begin{equation}
(D\exp(\bA))[\bB] = \D{}{\eps}\exp(\bA+\eps \bB) \Big|_{\eps=0} 
= \int_0^1 \er^{(1-\alpha) \bA} \bB \er^{\alpha \bA}\,d\alpha ~, 
\label{eq:Frechet3}
\end{equation}
where $\bA$ and $\bB$ are two arbitrary second-rank tensors. In our case, we have, for any $\bX$,
\begin{align}
\D{(\bPsi^{-1}\cdot\bBe)}{\bQ} \cdot \bX
& = \D{}{\eps}[\er^{-b(\bQ+\eps \bX)}\cdot \bBe ]\Big|_{\eps=0} \notag \\
& = - b \int_{0}^{1} \big(\er^{-(1-\alpha)\,b \bQ} \bX \er^{-\alpha\,b \bQ}\big)\cdot \bBe d\alpha \notag \\
& =  - b \bX \cdot \int_{0}^{1} \er^{-(1-\alpha)\,b \bQ} \bBe \er^{-\alpha\,b \bQ} d\alpha \notag \\
& =  - b \bX \cdot \int_{0}^{1} \er^{\alpha\,b \bQ} (\bPsi^{-1} \bBe) \er^{-\alpha\,b \bQ} d\alpha
\end{align}
where we have used $\bPsi^{-1} = \er^{-b \bQ}$ and the fact that $\bPsi^{-1}$ commutes with $\er^{\alpha\,b \bQ}$. Therefore the derivative of the neo-Hookean term with respect to $\bQ$ is
\begin{equation}
\D{}{\bQ}\,\tr\left(\bPsi^{-1}\bBe\right) 
	=  - b \int_{0}^{1} \er^{\alpha\,b \bQ} (\bPsi^{-1}\bBe) \er^{-\alpha\,b \bQ} d\alpha,
\label{eq:Frechet4}
\end{equation}
so that the evolution of the tensor $\bQ$ is governed by 
\begin{align}
\str{\bh} & = \rho \left( A \bQ - B \Big(\bQ^2 - \frac{1}{3}\tr(\bQ^2)\,\, \bI \Big) + C\,\tr(\bQ^2) \bQ \right) \notag 
\\
& -\frac{1}{2}\rho \mu b \int_{0}^{1} \str{\big[\er^{\alpha\,b \bQ} (\bPsi^{-1} \bBe) \er^{-\alpha\,b \bQ}\big]} 
d\alpha 
- \divr\left(\rho k  \nabla \bQ\right) = 0.
\label{eq:evoQ1}
\end{align}

A useful special case of Eq.\eqref{eq:Frechet3} that it is worth reporting is 
\begin{equation}
\big(\er^{b \bQ(\bx,t)}\big)\dot{\phantom{i}} = b \int_0^1 \er^{(1-\alpha)b \bQ} \dot{\bQ} \er^{\alpha b\bQ}\,d\alpha.
\label{eq:dot_eQ}
\end{equation}

We can try to simplify Eq.\eqref{eq:Frechet4} by using Eq.\eqref{eq:Quniaxial} and Cayley-Hamilton theorem to write 
\begin{align}
\er^{\alpha\,b \bQ} & = a_1(S,\alpha) \bQ + a_2(S,\alpha) \bQ^2 , \\
a_1(S,\alpha) & = \big(\tfrac{1}{2}\er^{2 \alpha  b S/3} - 2 e^{-\alpha b S/3}\big)/S ,\\
a_2(S,\alpha) & = 3\big(\tfrac{1}{2}\er^{2 \alpha  b S/3} + e^{-\alpha b S/3}\big)/S^2 \, .
\end{align}
When inserted in Eq.\eqref{eq:Frechet4}, the integral is then reduced to the calculations of the integrals of the 
products $a_1(S,\alpha)a_1(S,-\alpha)$, $a_1(S,\alpha)a_2(S,-\alpha)$, $a_2(S,\alpha)a_1(S,-\alpha)$, 
$a_2(S,\alpha)a_2(S,-\alpha)$. The result is most conveniently written in terms of the tensor $\bN=\bn \tp \bn - 
\tfrac{1}{3}\bI$. After some algebra, not reported for brevity, we find
\begin{subequations}
\begin{align}
\str{\left(\D{}{\bQ}\,\tr\left(\bPsi^{-1}\bBe\right) \right)}
& =  c_1\bN\bBe\bN
+ c_2 (\bN \bBe \bN^2 + \bN^2\bBe \bN) 
+ c_3 \bN^2 \bBe \bN^2 + \frac{b}{3}\tr\left(\bPsi^{-1}\bBe\right) \bI \\
c_1 & = -4 b \er^{b S/3}-\frac{1}{4} b \er^{-2 b S/3} + 2 (\er^{b S/3}-\er^{-2 b S/3})/S, \\
c_2 & = 6 b \er^{b S/3}-\frac{3}{4} b \er^{-2 b S/3} + 3(\er^{b S/3} - \er^{-2 b S/3})/(2S) , \\
c_3 & = -9 b \er^{b S/3}-\frac{9}{4} b \er^{-2 b S/3} - 9 (\er^{b S/3} - \er^{-2 b S/3})/S.
\end{align}
\label{eq:DQ_ver2}
\end{subequations}
Eq.\eqref{eq:evoQ1} then reads
\begin{align}
\str{\bh} & = \rho \left( A \bQ - B \Big(\bQ^2 - \frac{1}{3}\tr(\bQ^2)\,\, \bI \Big) + C\,\tr(\bQ^2) \bQ \right) \notag 
\\
& +\frac{1}{2}\rho \mu \Big( c_1\bN\bBe\bN
+ c_2 (\bN \bBe \bN^2 + \bN^2\bBe \bN) 
+ c_3 \bN^2 \bBe \bN^2 + \frac{b}{3}\tr\left(\bPsi^{-1}\bBe\right) \bI\Big) 
- \divr\left(\rho k  \nabla \bQ\right) = 0.
\end{align}

In many situations the material is nearly incompressible so that a good approximation the density $\rho$ is constant. 
Hence, we can write \eqref{eq:evoQ1} as
\begin{equation}
A \bQ - B \Big(\bQ^2 - \frac{1}{3}\tr(\bQ^2)\,\, 
\bI \Big) + C\,\tr(\bQ^2) \bQ 
-\frac{1}{2} \mu b \int_{0}^{1} \str{\big[\er^{\alpha\,b \bQ} (\bPsi^{-1} \bBe) \er^{-\alpha\,b \bQ}\big]} d\alpha
- k  \Delta \bQ = 0.
\label{eq:evoQ2}
\end{equation}

\section{Ordering tensor equation for the  active fluid approximation}
\label{app:calculations_fast}

When $b \ll 1$ we approximate the shape tensor as $\bL \approx \bI + b \bQ$ so that $\bL^{\uctd} \approx \bI^{\uctd} + 
b 
\bQ^{\uctd} = -2\bD + b \bQ^{\uctd}$. We can now use this approximation and Eq.\eqref{eq:LB_fast} to simplify the 
integral in Eq.\eqref{eq:moleculafieldQ}. Let us define $\bX := \bTa -\tau \bL^{\uctd}$ and consider its expansion in 
$b$: $\bX = \bX_0 + b \bX_1 + o(b) = \bTa +2\tau \bD - b\tau \bQ^{\uctd} + o(b)$. In agreement with the usual 
presentations in this context we introduce the co-rotational time derivative $\bQd := \dot{\bQ} - \bW\bQ+\bQ\bW$, with 
$\bW = \big(\nabla\bv - (\nabla\bv^T)\big)/2$, so that we can write the co-deformational time derivative as
\begin{equation}
\bQ^{\uctd} = \bQd - \bD\bQ-\bQ \bD. 
\end{equation}
Finally, the integral in Eq.\eqref{eq:moleculafieldQ} simplifies to
\begin{align}
\int_{0}^{1} \er^{\alpha\,b \bQ} (\bL^{-1} \bBe) \er^{-\alpha\,b \bQ} d\alpha 
& = \bI + \int_{0}^{1} \er^{\alpha\,b \bQ} \big(\bTa -\tau \bL^{\uctd}\big) \er^{(1-\alpha)\,b \bQ} d\alpha \notag \\
& = \bI + \int_{0}^{1} \big(\bI + \alpha b \bQ \big) 
\big(\bX_0 + b \bX_1 \big) \big(\bI + (1-\alpha) b \bQ\big) d\alpha \notag \\
& = \bI + \bTa +2\tau \bD +\frac{b}{2}\big(\bQ\bTa + \bTa\bQ\big) 
+ 2b\tau \big(\bQ\bD + \bD \bQ \big)
- b\tau \bQd. 
\label{eq:integral_fast}
\end{align}

%%%%%%%%%% Insert bibliography here %%%%%%%%%%%%%%
%\nocite{*}
%\bibliographystyle{plain}
%\bibliographystyle{siam}
\bibliographystyle{unsrt}
\bibliography{RefActiveQ.bib}

\end{document}